\documentclass[journal]{IEEEtran}

\usepackage{enumitem,calc}
\usepackage{amsmath,graphicx,cite}
\usepackage[ruled,linesnumbered]{algorithm2e}
\usepackage{cite,graphicx,epsfig,wrapfig,amsfonts,amssymb,threeparttable,color,url}
\usepackage{amsthm}
\usepackage{subfigure}
\usepackage{epsf}
\usepackage{booktabs}
\usepackage{algpseudocode}
\usepackage{float}
\usepackage{bm}
\usepackage{footnote}
\usepackage{fixmath}

\theoremstyle{definition}

\theoremstyle{remark}

\begin{document}
	
\title{Delay-aware Resource Allocation in Fog-assisted IoT Networks Through Reinforcement Learning}
\author{
\IEEEauthorblockN{Qiang Fan,~\IEEEmembership{Member,~IEEE}, Jianan Bai, Hongxia Zhang, Yang Yi, and Lingjia Liu,~\IEEEmembership{Senior Member,~IEEE} \thanks{Q. Fan, J. Bai, Y. Yi, and L. Liu are with Department of Electrical and Computer Engineering, Virginia Tech, Blacksburg, VA, 24060 USA. H. Zhang is with College of Computer Science and Technology, China University of Petroleum, Qiangdao, 266580, China. The work of H. Zhang was completed when she was visiting Virginia Tech.
}
\thanks{Q. Fan, J. Bai, Y. Yi, and L. Liu are partially supported by U.S. National Science Foundation (NSF) under grants ECCS-1811497 and CCF-1937487. The corresponding author is L. Liu (ljliu@ieee.org).}
}
}
\maketitle

	\begin{abstract}
		Fog nodes in the vicinity of IoT devices are promising to provision low latency services by offloading tasks from IoT devices to them. Mobile IoT is composed by mobile IoT devices such as vehicles, wearable devices and smartphones. Owing to the time-varying channel conditions, traffic loads and computing loads, it is challenging to improve the quality of service (QoS) of mobile IoT devices. As task delay consists of both the transmission delay and computing delay, we investigate the resource allocation (i.e., including both radio resource and computation resource) in both the wireless channel and fog node to minimize the delay of all tasks while their QoS constraints are satisfied. We formulate the resource allocation problem into an integer non-linear problem, where both the radio resource and computation resource are taken into account. As IoT tasks are dynamic, the resource allocation for different tasks are coupled with each other and the future information is impractical to be obtained. Therefore, we design an on-line reinforcement learning algorithm to make the sub-optimal decision in real time based on the system's experience replay data. The performance of the designed algorithm has been demonstrated by extensive simulation results.  
		

	\end{abstract}
	
	\section{Introduction}
	\IEEEPARstart{R}{e}cently, a tremendous number of mobile smart devices, such as autonomous vehicles, wearable devices and smartphones have been extensively employed in people's daily life. These devices enable various IoT applications, such as autonomous driving, smart health, smart city and smart home. Owing to the high volume and fast velocity of data streams generated by mobile IoT devices, the cloud can be utilized to provision flexible computation and storage resources for these IoT devices \cite{wang2015processing}. However, since the data source is far away from the cloud and  the data streams have to go through the Internet before being transmitted to the cloud, the transmission delay of IoT tasks may be unbearable for some delay sensitive applications such as autonomous driving and augmented reality \cite{Gunaseelan2010performance}. To tackle this problem, the fog computing is introduced to place computation resources at gateways and thus processes IoT tasks at the network edge, which significantly reduces the transmission delay of IoT tasks \cite{fan2018workload, fan2019on}. Due to the complex network, intelligent fog network leveraging machine learning methods (i.e., consisting both deep learning and reinforcement learning) \cite{wang2017automatic,Dinh2018learning,wang2018segmentation} is promising to learn the network features  and thus effectively manage the network resources.  
	
	In fog-assisted mobile IoT networks, the task delay consists of both the wireless transmission delay and computing delay and thus is impacted by the resource allocation in both the wireless channel and fog node. As tasks are generated dynamically, the optimal decision on radio resource allocation requires the complete network information such as available bandwidth, channel conditions of IoT devices, and the traffic sizes of all tasks. The real time radio resource allocation for different IoT tasks are coupled with each other owing to the limited bandwidth of the system\cite{Li2015resource}. Specifically, more bandwidth allocated for the current task deprives the bandwidth for the following tasks. However, it is challenging to obtain the future information such as the channel condition and task information in advance. In this case, optimizing the radio resource allocation based on the complete network information is impossible, and thus the online algorithm based on the current network information in absence of further information is required to obtain the sub-optimal solution. Similarly, at the side of a fog node, the computation resource allocated for current task will also affect the computation resources for future tasks. Meanwhile, due to the quality of service (QoS) requirement of IoT tasks (i.e., in terms of maximum allowed task delay), the radio resource allocation and computation resource allocation are coupled with each other for each task. In other words, a task allocated with more bandwidth owing to its desirable channel condition can be provisioned with less computation resource, thus saving computation resources for other devices with poor channel conditions. Currently, most existing works focus on allocating radio resources or computing resources to improve the network performance based on the given network information. However, few works have paid attention on the joint radio and computing resource allocations for different tasks by reinforcement learning to adapt to the dynamic network condition. Since the task arrivals are dynamic and the available resources in the system is time-varying, it is challenging to dynamically allocate both the radio and computing resources to various tasks in real-time to improve the delay of each IoT task, where the future task information is impractical to be obtained. This is the main motivation of this article.  
	
	To solve this problem, we propose a delay-aware online resource allocation algorithm based on reinforcement learning to allocate radio and computation resources for IoT tasks to reduce their task delay. Our contribution can be summarized as follows:
	
	\begin{itemize}[leftmargin=*]
		\item We investigate the joint radio and computation resource allocation problem to reduce the task delay in fog-assisted mobile IoT networks with incomplete network information. We mathematically formulate the problem as an integer non-linear problem that is challenging to solve.  In the formulation, we have considered the basic granularity of both radio and computation resources to make it applicable in practical engineering. 

	\item We have considered the delay requirements of all tasks such that each take the meet the QoS requirement in the joint radio and computation resource allocation.

	\item To efficiently solve this problem, we design an online resource allocation algorithm based on reinforcement learning (i.e., actor-critic), in which an agent can learn from the dynamic environment to make desirable decisions. We have conducted extensive simulations to demonstrate the performance of our algorithm.

	\end{itemize}
	The remainder of this paper is organized as follows. In Section \ref{secRelated}, we briefly review related works. In Section \ref{secSystem}, we illustrate the fog-assisted mobile IoT network and introduce the system model. In Section \ref{secProblem}, we formulate and analyze the resource allocation problem for IoT tasks. In Section \ref{secThe}, the resource allocation algorithm based on reinforcement learning is proposed to obtain the suboptimal solution of the above problem. Section \ref{secNumerical} shows the simulation results, and concluding remarks are presented in Section \ref{secConclusion}. 

	\section{Related Works}
	\label{secRelated}
	Fog computing is promising to provide low latency service for IoT tasks, owing to its proximity to IoT devices. As workload distribution in the network is spatially and temporally dynamic, some studies have focused on workload allocation in fog computing, especially for delay sensitive application such as autonomous vehicle, augmented reality \cite{deng2016optimal,fan2018towards,wan2018fog,wu2020a}. Zeng \textit{et al.} \cite{zeng2016joint} jointly optimized the task scheduling and image placement to improve the task delay in fog networks. Fan and Ansari \cite{fan2018workload} designed a workload allocation scheme based on the different cloudlet capacities in a hierarchical cloudlet network to minimize task delay, where the wireless transmission delay is neglected. Jia \textit{et al.} \cite{jia2015optimal} investigated to place cloudlets in the network and balance the workload among distributed cloudlets and thus reduce the task delay, where the radio resource allocation is ignored. Fan \textit{et al.} \cite{fan2017energy} investigated to migrate virtual machines from green energy deprived cloudlets to green energy overprovisioning cloudlets to fully utilize the green energy in the network. However, all these works emphasize utilizing workload allocation among edge servers to enhance the user experience or energy efficiency of the network instead of focusing on resource allocation \cite{abedin2019resource,yu2018uplink}. 
	
	Some researchers also considered the computation resource allocation or radio resource allocation in fog-assisted IoT network to further enhance the network performance \cite{Liu2007resource, Liu2014energy}. Tong \textit{et al.} \cite{tong2016cloud} investigated the cloudlet selection and computation resource allocation for tasks in hierarchical cloudlet network, instead of radio resource allocation. However, they assumed that all the tasks' information is ready at the beginning of each slot and only optimized the computation resources for different tasks without considering the dynamic network condition. Fan \textit {et al.} \cite{fan2018application} proposed to offload each application's workloads among different cloudlets and allocate computation resources of each cloudlet to different types of tasks based on its workload; however, they neglected the radio resource allocation and emphasized the long term performance instead of the real-time performance. Tran \textit{et al.} \cite{tran2019joint} proposed a task offloading and resource allocation scheme in mobile edge computing to maximize the offloading gains in terms of both the delay reduction and energy reduction. In this work, the joint problem is decomposed into two subproblems, and thus the authors make the task offloading decision and allocate computation resource of each edge server for user tasks. Lyu \textit{et al.} \cite{lyu2017multiuser} proposed a heuristic algorithm to allocate computation resources to the offloaded tasks. Since each user accesses one wireless channel, they only considered the computation resource allocation without considering the mutual effect between radio resource and computation resource allocations. In addition, other researchers emphasized the radio resource allocation instead. Dab \textit {et al.}\cite{dab2019joint} designed a new joint task assignment and radio resource allocation scheme in the WiFi-based mobile edge computing. The objective of the work is to reduce the energy consumption of users while satisfying QoS requirement. However, they focused on the communications issues without considering the impact of computation resource allocation on the network performance. Zhao \textit{et al.} \cite{zhao2019deep} employed multi-agent reinforcement learning algorithm to jointly associate users to base stations and allocate channels to users thus achieving the maximum network utility. All the above works have not considered to employ the joint radio and computation resource allocation to improve the performance of each IoT task in real-time without the complete network information. Wang \textit{et al.} \cite{Wang2019task} proposed to allocate the transmit power and wireless channel to improve the execution delay of tasks by reinforcement learning methods. Alfakih \textit{et al.} \cite{Alfakih2020task} allocated tasks among various edge servers and then focused on bandwidth allocation for different tasks based on reinforcement learning methods to minimize the system cost. However, they only emphasized the communications sector while assuming the computing resources for various tasks are fixed. Liu \textit{et al.} \cite{Liu2019resource} introduced a reinforcement learning algorithm to determine the task offloading and power allocation for mobile user in order to optimize the tradeoff between power consumption and the task delay. However, they just neglected the bandwidth and computation resource allocation that crucially impact the execution delay of tasks. 
	
	Most existing works assume that the workload of the network is given in advance, and optimize the network performance (e.g., the average task delay within a long period) based on the global network information. However, as the future task information and network status are usually hard to predict, it is impractical to allocate optimal resources for the arriving tasks in real time based on the global information. On the other hand, IoT task delay is impacted by both the radio and computation resource allocation as it is composed of both the transmission delay and computing delay. However, few works have paid attention on the radio resource allocation and computation resource allocation simultaneously in real time, and this issue remains an open challenge. Therefore, we propose an online resource allocation algorithm to enhance the task delay, where both the radio resource and computation resource are taken into account. In our scheme, the resource allocations of the current task and future task are coupled with each other while the radio resource allocation is related to the computation resource allocation for an individual task. Different from other papers that continuously allocate radio or computation resource to tasks, we also consider the granularity of these resources to make it more applicable to the realistic network. As the wireless channel of a mobile IoT device is time-varying as well as the fog node status, the resource allocation decision should be determined based on different wireless channel conditions, fog node status and task information.                 
	
	\section{System Model}
	\label{secSystem}
	\begin{figure}[!htb]
		\centering	
		\includegraphics[width=1\columnwidth]{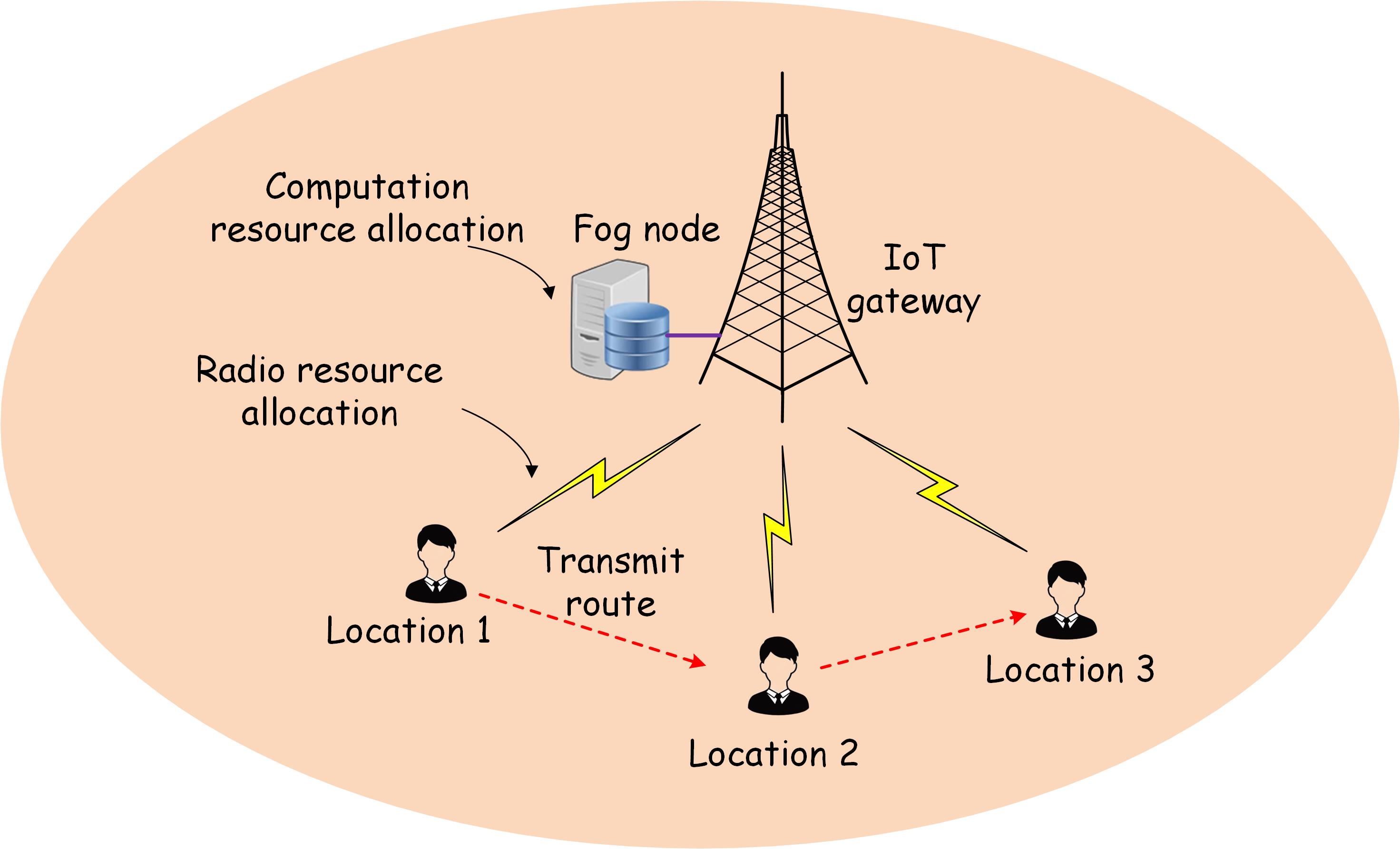}
		\caption{Fog-assisted IoT network.}
		\label{fig_framework}
	\end{figure}
	
	A fog-assisted mobile IoT network has been illustrated in Fig. 1. In this paper, we employ the cellular network as our IoT network infrastructure with base stations (BSs) acting as IoT gateways (GWs) to provision communications service for IoT devices. Each GW is equipped with a fog node to provide computation and storage resources at the network edge \cite{fan2019towards}. The fog node is responsible for making resource allocation decisions in real time based on the network status. The GW can detect the wireless conditions towards IoT devices and send them to the fog node. Based on this fog-assisted mobile IoT network, tasks of IoT devices can be transferred to their GW and processed by the corresponding fog node. Generally, each mobile IoT device may visit several locations based on a certain route. At each location, it collects data and transfers IoT tasks to the fog node for processing \cite{yao2019task}.  Owing to the mobility of IoT devices, their channel conditions are time-varying. Meanwhile, as IoT tasks are generated at different time, the fog node status keeps changing with the time-varying workload. As the task delay consists of both the transmission delay and computing delay, it is impacted by both the radio and computation resource allocation in the network. If an IoT device has bad channel condition while the available bandwidth is insufficient, it requires more computation resource to ensure low task delay; otherwise, it can be allocated with more bandwidth while saving computation resource for other IoT devices. Note that both the radio and computation resource cannot be continuously allocated for tasks in practical engineering, and thus the basic granularity of radio and computation resource are denoted as $\alpha$ (in Hz) \cite{kotagi2016breathe} and $\beta$ (in CPU cycle/s) \cite{tong2016cloud}, respectively. Accordingly, we define a resource block as the granularity of radio resource (i.e., $\alpha$) and a computation unit as the granularity of computation resource (i.e., $\beta$). 
	
	In this paper, we denote $\mathcal{I}$ as the set of all IoT tasks and $i$ as the index of an IoT task within $\mathcal{I}$. Denote $x_{i}$ as the number of resource blocks allocated to task $i$, and $y_i$ as the number of computation units allocated to task $i$. Hence, the radio and computation resource allocated to task $i$ becomes $x_{i}\alpha$ and $y_i \beta$, respectively. The key notations used in this paper are listed in Table \ref{tab_network}.
	\begin{table}[!htb]
	\renewcommand{\arraystretch}{1.3}
	\caption{Key notations}
	\label{tab_network}
	\centering
	\begin{tabular}{cl}
		\toprule[1.5pt]
		
		\textbf{Symbol} & \textbf{Definition}   \ \ \ \ \  \\
		\hline
		Number of resource blocks for a task     & $x_i$ \\
		Number of computation units for a task    & $y_i$ \\
		Computation intensity of task $i$ & $\mu_i$ CPU cycle/bit   \\
		Data size of task $i$	&	$l_i$ bits		\\
		Computation size of task $i$                & $c_i$ CPU cycles\\
		Resource block 		&  $\alpha$ Hz   \\
		Computation unit		&	$\beta$ CPU cycles/s 		\\
		Task delay of task $i$   & $D_i$  \\
		Transmission delay of task $i$ & $D^t_i$  \\
		Computing delay of task $i$  & $D^c_i$ \\
		QoS requirement  & $\tau_0$ \\
		Maximum number of resource blocks of the system & $M$ \\
		Maximum number of computation units of the system& $N$ \\
		\bottomrule[1.5pt] 
		
	\end{tabular}
\end{table}

	\subsection{Transmission Delay}
	 In order to process IoT tasks at a fog node, an IoT device has to transmit its tasks to the GW via uplink communications. The wireless uplink rate is mainly dependent on the wireless channel condition and the allocated radio resource. After the fog node processes a task, it needs to feedback the processing results to the corresponding IoT device. However, since the processing results are much smaller than IoT tasks and have high data rate in the wireless downlink channel, the downlink delays of the results have been neglected \cite{fan2019towards}. In this paper, we just focus on the uplink communications of IoT devices.
	 
	 Denote $P_i$ as the transmission power of the IoT device with task $i$, $h_{i}$ as the channel gain between the IoT device and GW, $\sigma^2$ as the noise power. The frequency efficiency of the IoT device can be derived according to the Shannon Hartley theorem \cite{fan2016green} as follows:
	\begin{equation}
	{\eta_{i}} = \log (\frac{{{P_i}{h_{i}}}}{{{\sigma ^2}}}).	
	\end{equation}                                                                      
	Hence, as the allocated radio resource is $x_{i}\alpha$, the uplink data rate can be expressed as
	\begin{equation}
	    r_{i}=x_{i}\alpha\eta_{i}.
	\end{equation}
	Given the data size of task $i$, the transmission delay of task $i$ can be expressed as
	\begin{equation}
	D^t_i=\frac{l_i}{x_{i}\alpha\eta_{i}}.
	\end{equation} 
	
	
	\subsection{Computing Delay}
	The computing delay of task $i$ depends on the allocated computation resource and the computation size of task $i$. As the computation intensity of task $i$ is denoted as $\mu_i$ (CPU cycle/ bit), the computation size of task $i$ is a function of its data size and can be expressed as $c_i=\mu_i l_i$. Therefore, the computing delay of task $i$ can be derived as 
	\begin{equation}
	D^c_i=\frac{c_i}{y_{i}\beta}.
	\end{equation}  
	
	Aggregating both the transmission delay and computing delay, we can derive the task delay of task $i$ as
	\begin{equation}
	D_i=D^t_i+D^c_i.
	\end{equation}
	
	\section{Problem Formulation}
	\label{secProblem}
	
	  The task delay will be affected by different factors, such as channel condition, the available radio and communications resource of the network, and computation intensity. First, if a task has bad channel condition, it is preferable to be allocated with less radio resource, and thus more radio resources can be allocated to other tasks with the desirable channel conditions. Therefore, the high spectrum efficiency of the network will significantly improve the task delay of all tasks. Second, the resources (i.e., either radio or computation resource) allocated for different tasks are coupled with each other. For example, if task A obtains a large number of resource blocks, the system may not have sufficient resource blocks for the following task B even if task B has better channel conditions than task A. Third, if the remaining radio resource is insufficient and incurs a high transmission delay for a task, the fog node is forced to allocated more computation resources to the task to meet the QoS requirement. Forth, the computation resource allocation is also impacted by the heterogeneous computation intensities of tasks. Note that the data size and computing size of the tasks significantly impact the QoS of tasks and thus we focus on the workload jobs in this paper. The main goal of this paper is to minimize the task delay of IoT tasks offloaded by IoT devices, while satisfying the QoS requirement of each task. Thus, we can formulate the resource allocation problem as follows:     
	
	\begin{align}
	P1: \mathop {\min }\limits_{{x_{i},y_{i}}}    &\frac{1}{\lvert \mathcal{I} \rvert}\sum\limits_{i\in \mathcal{I}} { {\frac{{{l_i}}}{{{x_{i}}\alpha{\eta_j}}} + \frac{{{c_i}}}{{{y_{i}}\beta }}} } \\
	s.t.,   &\frac{{{l_i}}}{{{x_{i}}{\eta_j}}} + \frac{{{c_i}}}{{{y_{i}}\beta }} <  = {\tau _0},\ \forall i\in \mathcal{I}, \label{Const_1}\\
	&\sum\limits_i { {{x_{i}}}  <  = M},\ \forall i\in \mathcal{I} \label{Const_2}\\
	&\sum\limits_i { {{y_{i}}}  <  = N},\ \forall i\in \mathcal{I}. \label{Const_3}
	\end{align}	
	Here, $\tau_0$ is the QoS requirement of a task in terms of maximum allowed task delay. Constraint (\ref{Const_1}) ensures each task to satisfy the QoS requirement. Constraint (\ref{Const_2}) imposes that the total utilized resource blocks to be no more than the maximum number of resource blocks of the system. Constraint (\ref{Const_3}) imposes the total utilized computation resources to be no more than the capacity of a fog node.  
	
	Optimizing the resource allocation requires the complete task information. However, the complete future task information is difficult to predict in advance, and thus it is impractical to obtain the optimal solution with the existing network status. On the other hand, even if the complete task information is provided, the above problem is an integer non-linear problem and thus is challenging to solve.  To obtain the optimal resource allocation decision, a brute-force search leads to $O(M^{\lvert \mathcal{I} \rvert}N^{\lvert \mathcal{I} \rvert})$ iterations where $\lvert \mathcal{I} \rvert$ represents the total number of tasks. The computational complexity of the brute-force search increases exponentially with respect to the total number of tasks. Hence, optimizing the resource allocation in real time becomes impractical, especially for a large-scale network \cite{fan2018throughput}.

\section{The Resource Allocation Algorithm}
\label{secThe}

Due to the unawareness of future task information and high complexity of P1, we hence design an Online Resource Allocation algorithm (ORA) based on reinforcement learning to efficiently solve the above problem in real time. Essentially, ORA learns the environment over many epochs, in each of which it takes actions for many steps (i.e., for task arrivals) to maximize the reward of the system. 

In the network, the amount of available radio and computation resource is impacted by different events such as the arrival and departure of an IoT task. When an IoT task arrives, the system has to make a decision to allocate both the radio resource and computation resource to process the task. Meanwhile, when the task departs the system after task processing, the system just updates the available resources accordingly without making any decision \cite{wu2020delay}. Through the resource allocation decision, the system can significantly improve a reward that depends on the QoS of tasks. 

To employ the reinforcement learning to solve P1, the resource allocation problem is further represented as a four-dimensional tuple $(\mathcal{S}, \mathcal{A}, \mathcal{T}, \mathcal{R})$, where $\mathcal{S}$ is the set of all possible states, $\mathcal{A}$ is the set of all possible actions, $\mathcal{T}: \mathcal{S}\times\mathcal{A}\rightarrow\mathcal{S}$ is the state transition function mapping from a state and an action to the next state, and $\mathcal{R}: \mathcal{S}\times\mathcal{A}\rightarrow \mathbb{R}$ is the reward function measuring the benefit of selecting an specific action under a given state \cite{li2019reinforcement}. 

In this paper, a state stands for the set consisting of the remaining radio resource, the remaining computation resource, data size of the arriving task, and the computation size of the arriving task. Once a task arrives, the action of an agent reflects both the radio resource and computation resource allocated to the task, and thus is defined as joint action. Note that the state and joint action are denoted as $s$ and $a$, respectively. Since the goal of this paper is to minimize the task delay, the reward of the joint action is defined as $r=-D_i$, where $D_i$ is the delay of the task. Essentially, with the arrival of a task, we need to select a joint action based on current state, and thus enhance the reward of the system.         

In ORA, the fog node serves as an agent that iteratively learns to make a right decision to react to the current state, i.e., trying to find an optimal policy, $\pi: \mathcal{S}\rightarrow\mathcal{A}$, in terms of maximizing a discounted future reward $R = \sum_{t=0}^T\gamma^tr_t$, where $T$ is the time horizon, $r_t$ is the immediate reward at time $t$, and $\gamma\in[0, 1]$ is a discount factor. In this paper, due to the large action space of the joint action $(x,y)$, we employ the actor-critic approach of reinforcement learning with high computational efficiency to achieve the policy \cite{tathe2018dynamic}, where the agent is equipped with two neural networks: actor network and critical network. Note that the actor-critic approach is a combination of Q-learning algorithm and policy gradient algorithm. 

\subsubsection{Q-Learning}
Q-learning is a family of value-based reinforcement learning algorithms, which estimate the action-value function $Q^\pi(s,a) = \mathbb{E}[R_t|s_t = s, a_t = a]$ under the policy $\pi$. The action-value function can be derived through the well-known Bellman function $Q^\pi(s,a) = \mathbb{E}_{s'}[r(s, a) + \gamma\mathbb{E}_{a'}[Q^\pi(s', a')]]$. Since the exact form of action-value function can be extremely difficult to obtain in practice, we generally parameterize $Q^\pi(s, a)$ as $Q^\pi(s, a;\theta)$ using a deep neural network, where $\theta$ is the network parameters. The action-value function $Q^*$ corresponding to the optimal policy can be obtained by minimizing the loss
\begin{equation}
    L(\theta) = \mathbb{E}_{(s, a, r, s')\sim\mathcal{D}}[(y - Q(s, a; \theta))^2],
\end{equation}
where $y = r + \gamma\max_{a'}Q(s', a';\theta)$ and $\mathcal{D}$ is the experience buffer. The optimal policy can be written as $\pi^* = \arg \max_aQ^*(s,a)$.

\subsubsection{Policy Gradient}
Differing from value-based reinforcement learning paradigm, policy gradient algorithms directly parameterize the policy as $\pi_\theta(a|s)$, which represents the probability of choosing action $a$ under a given state $s$. The parameter $\theta$ is updated to maximize the objective $J(\theta) = \mathbb{E}_{s, a}[\pi_\theta(a|s)q(s,a)]$, where $q(s,a)$ is a value function to measure how good the action $a$ is. Then, the policy can be optimized by adjusting the parameters $\theta$ along the direction of policy gradient
\begin{equation}
    \nabla_\theta J(\theta) = \mathbb{E}_{s, a}[\nabla_\theta \ln\pi_\theta(a|s)q(s,a)].
\end{equation}
Different definitions of $q(s,a)$ lead to different algorithms. For example, REINFORCE algorithm simply uses a sample return $\sum_{i=t}^T\gamma^{i-t}r_i$ as the value function. On the other hand, using the action-value function $Q^\pi(s,a)$ defined for Q-learning as the value function results in \textit{actor-critic} algorithms, which have the advantage of ameliorating variance during training. In practice, the action-value function $Q^\pi(s,a)$ is usually replaced by an advantage function $A^\pi(s,a) = Q^\pi(s,a) - b(s)$, where $b(s)$ is a state-related baseline to further mitigate variance and accelerate training. Actor-critic algorithms combines the merits of Q-learning and policy gradient, and it is very popular in recent years.

\subsubsection{Actor-critic}
By combining Q-learning with policy gradient, we employ actor-critic to allocate radio and computation resources for tasks in real time. Specifically, in actor-critic, an agent is equipped with two neural networks, namely \textit{actor network} and \textit{critic network}. 
When a task is generated at the mobile device, the actor network takes the state input $s = (e, c, d, l)$, where $e$ is the number of remaining resource blocks, $c$ is the number of remaining computation units, $d$ is the data size, and $l$ is the computation size. By forwarding the state $s$, the actor network outputs two category distributions $p_\theta(x|s)$ and $p_\theta(y|s)$, where $x=0,1,\cdots, M$, $y=0,1,\cdots, N$, and $\theta$ is the parameters of the actor network. The policy is then denoted as $\pi_\theta(a|s) = p_\theta(x|s)p_\theta(y|s)$, which gives the probability of choosing the joint action $a$. According to the two distributions, the actor selects a joint action $a = (x,y)$, where $x$ is the number of allocated radio resource blocks, and $y$ is the number of allocated computation units. The corresponding reward of the joint action $a$ is given by $r = -D_i$, where $D_i$ is the task delay of the task. The critic network takes the state $s$ as input and generates a state-value $V_w(s)$, where $w$ is the network parameter, to estimate the expected future reward starting from state $s$. Then, an advantage can be calculated as $A(s, a) = r - V_w(s)$, which measures how the joint action $a$ performs compared to our expectation.

The actor is trying to select an joint action $a$ with larger expected advantage, so that it updates the network parameters to maximize
\begin{equation}
    J(\theta) = \mathbb{E}_{(s,a)\sim\mathcal{D}}[\pi_\theta(a|s)A(s,a)],
\end{equation}
which results in the gradient direction
\begin{equation}
\label{eq: actor update}
    \nabla J(\theta) = \mathbb{E}_{(s,a)\sim\mathcal{D}}[\nabla_\theta\ln{\pi_\theta(a|s)}A(s,a)].
\end{equation}
To estimate a more accurate state-value, the critic will minimize the Euclidean norm between $V(s)$ and $r + V(s')$, and it leads to the gradient direction
\begin{equation}
\label{eq: critic update}
    \nabla_wL(w) = \mathbb{E}_{(s,r,s')\sim\mathcal{D}}[(r + V(s') - V(s))\nabla_wV(s)].
\end{equation}
The actor network and the critic network will be updated alternatively to maximize the expected future reward. We will update the two networks in each epoch until the predefined number of epochs is reached.

The flow of the proposed algorithm is shown as follows: In each epoch, when a task arrives, we can calculate the probabilities of different actions based on current state $s$. According the probabilities, the desirable action is selected. Then, the reward $r$ and the next state $s^{'}$ due to $a$ is derived, and thus the transition information, i.e. $(s,a,r,s^{'})$, is stored in memory $\cal D$. The above procedure is repeated for all tasks in the epoch. Afterwards, we retrieve transitions information with bath size $N$ from the memory, based on which the two neural networks for the actor and critic are trained by Eq. (13) and (14) respectively, i.e., achieving the corresponding parameters $\theta$ and $w$. We will repeat the above procedures for all epochs and derive the stable model parameters for the two neural networks. The detailed procedures are further shown in Algorithm \ref{alg_1}. 

\begin{algorithm}
	\caption{ORA Algorithm}\label{alg_1}
	\For {each training epoch}{
	\For {each arriving task}{
	 Forward the state input $s = (e, c, d, l)$ in the actor network to generate two Categorical distributions $p_\theta(x|s)$ and $p_\theta(y|s)$, for $x\in\mathcal{X}$ and $y\in\mathcal{Y}$\;
	 Sample and execute the action $a = (x, y)$ from the distributions $p_\theta(x|s)$ and $p_\theta(y|s)$\;
	 Observe the reward $r$ and the next state $s'$\;
	 Store the transition $(s, a, r, s')$ in memory $\mathcal{D}$\;}
	 Sample transitions from $\mathcal{D}$ with batch size $N$\;
	 Train the actor network and the critic network using gradients obtained by \eqref{eq: actor update} and \eqref{eq: critic update}\;}
\end{algorithm}

\textbf{Computational complexity.} We further analyze the complexity of the designed algorithm. The number of iterations (form Line 1 to Line 10) is determined by the number of epochs (denoted as $H$). The loop from Line 3 to Line 7 are executed for $\lvert \mathcal{I} \rvert$ times (i.e., equal to the number of IoT tasks), where the complexity of each time can be expressed as $O(2(M+N) )$. In addition, the complexity of Line 9 is related to the batch size and thus can be expressed as $O(\lvert \mathcal{D} \rvert)$. Therefore, the designed algorithm yields a computational complexity of $O(H*(2 \lvert \mathcal{I} \rvert (M+N)+\lvert \mathcal{D} \rvert))=O(H\lvert \mathcal{I} \rvert M+H\lvert \mathcal{I} \rvert N+H\lvert \mathcal{D} \rvert)$, and thus can achieve a solution in polynomial time.

\section{Numerical Results} 	
\label{secNumerical} 
In this section, we have set up simulations to verify the performance of the designed algorithm. To further validate the performance of the designed ORA algorithm, we also select two existing algorithms as baselines: Computation-only and Transmission-only. We utilize the Computation-only algorithm inspired by \cite{lyu2017multiuser} for comparison, which focuses on the computation resource allocation based on reinforcement learning, while the radio resource of the system is averagely allocated to tasks in each second, i.e., each task has the same radio resource. Meanwhile, Transmission-only \cite{Alfakih2020task} focuses on the radio resource allocation by reinforcement learning, while the total computation resource of the system is averagely allocated to tasks in one second. 

In the simulation, we consider an area of 1 $km^2$, i.e, the coverage area of a GW. There are 50 locations uniformly distributed in the network, where mobile IoT devices visit and offload IoT tasks to the fog node for task processing. Note that each mobile IoT device may select 5 locations and visit them, where the user mobility pattern does not affect the problem \cite{pomper2017design} since we just consider that the IoT device offloads tasks when stopping at a location. The total number of tasks over all locations is 500, and they are randomly generated among these 50 locations within a time duration of 50s. For the channel model, we employ the wireless path loss model, i.e., 128.1+37.6$log_{10}d$ from 3GPP specification \cite{wiley2011lte}, where $d$ is the distance in km. The data sizes of tasks are chosen according to the Normal distribution with an average of 1 Mbits and a variance of 0.3 Mbits, i.e., $N(10^6, 3*10^5)$. The computation intensity for different tasks is chosen based on $N(10, 3)$ (CPU cycle/bit). The QoS requirement is 1 s. Note that if the system does not have enough available resources for a task to satisfy the QoS requirement, we assume the task is dropped and the corresponding task delay is set to be 10 s. The remaining parameters are summarized as Table \ref{tab:para}.
\begin{table}[!htb]
	\renewcommand{\arraystretch}{1.3}
	\caption{Simulation Parameters}
	\label{tab:para}
	\centering
	\begin{tabular}{cl}
		\toprule[1.5pt]
		
		\textbf{Symbol} & \textbf{Definition}   \ \ \ \ \  \\
		\hline
		Number of IoT tasks		& 	500\\
		Data sizes of tasks		&	$N(10^6, 3*10^5)$ bits.		\\
		Computation intensity of tasks & $N(10, 3)$ CPU cycle/bit   \\
		Computation capacity of a fog		&	$3*10^8$ CPU cycle/s		\\
		System bandwidth                & 5 MHz   \\
		Radio resource block			& 180 kHz \\
		Computation granularity					& $1.0*10^7$ CPU cycle/s \\
		Transmission power of IoT device        & 200 mW    \\
		Noise power 							& -104 dBm	\\
		Path loss model							& $128.1+37.6{log}_{10} d$ ($d$ in  km)\\ 	
		QoS constraint 				     	& 1 s \\		
		\bottomrule[1.5pt] 
		
	\end{tabular}
\end{table}

	\begin{figure}[!htb]
		\centering	
		\includegraphics[width=1\columnwidth]{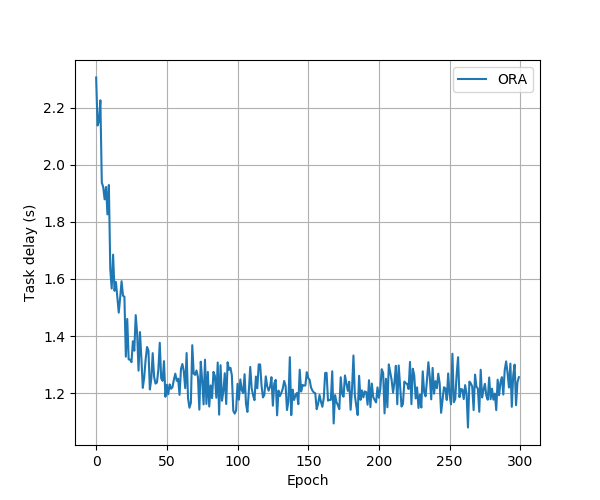}
		\caption{Learning process of the designed algorithm.}
		\label{fig_1}
	\end{figure}
	
	\begin{figure}[!htb]
		\centering	
		\includegraphics[width=1\columnwidth]{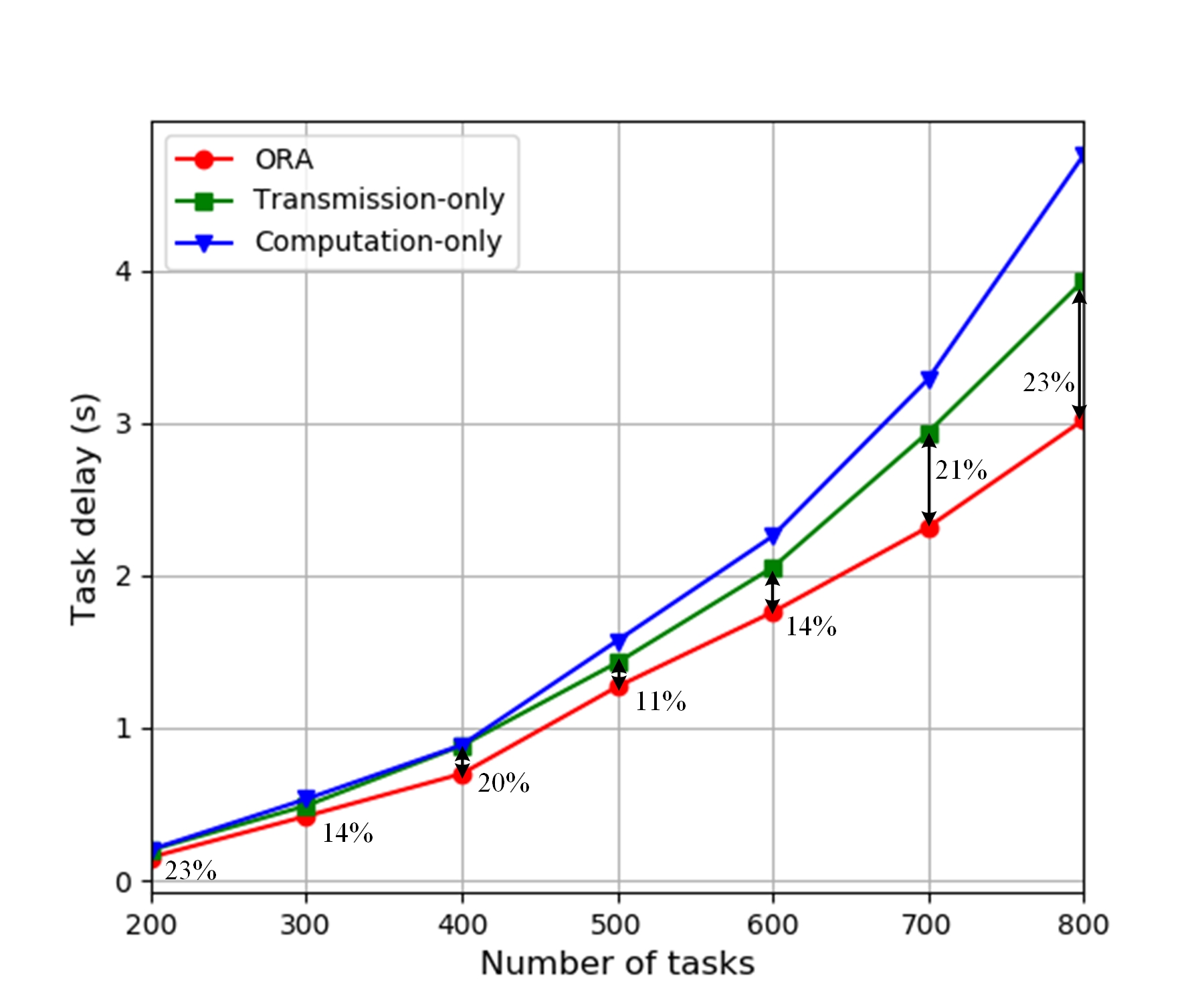}
		\caption{Average task delay with respect to different number of tasks}
		\label{fig_2}
	\end{figure}
	
	\begin{figure}[!htb]
		\centering	
		\includegraphics[width=1\columnwidth]{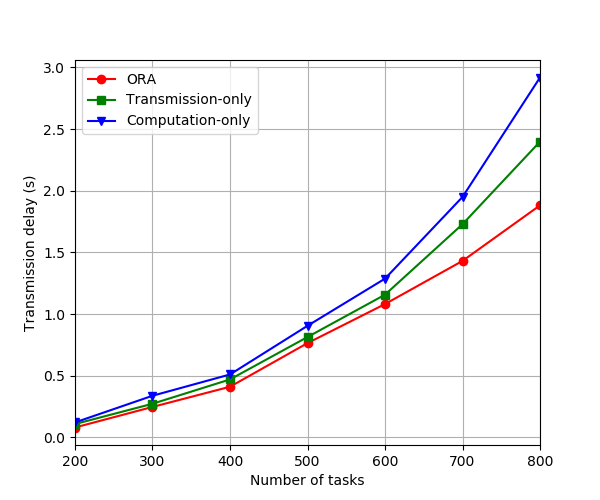}
		\caption{Transmission delay with respect to different numbers of tasks.}
		\label{fig_3}
	\end{figure}
	
		\begin{figure}[!htb]
		\centering	
		\includegraphics[width=1\columnwidth]{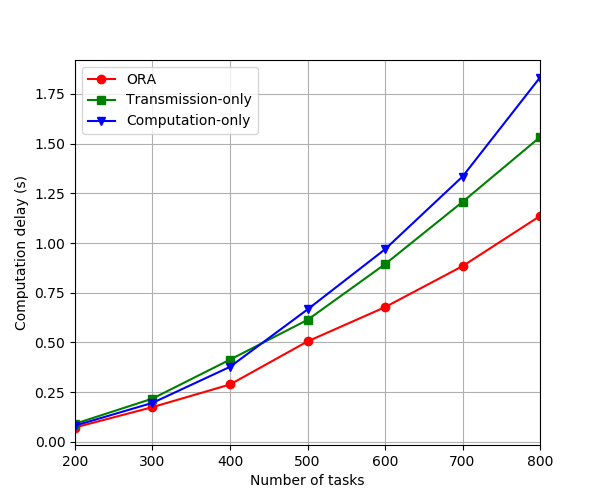}
		\caption{Computation delay with respect to different numbers of tasks.}
		\label{fig_4}
	\end{figure}
    \begin{figure}[!htb]
		\centering	
		\includegraphics[width=1\columnwidth]{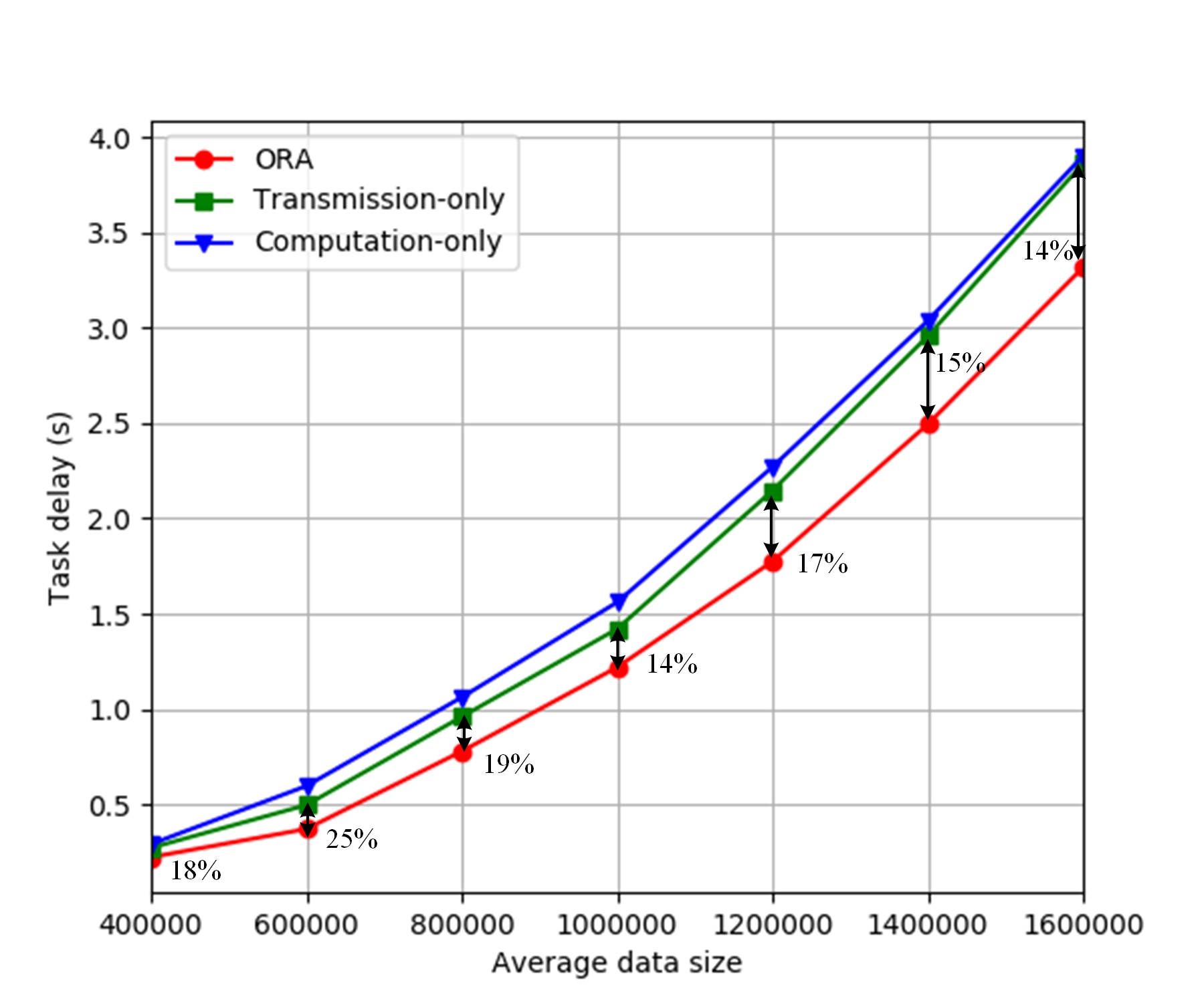}
		\caption{Task delay vs. average data size.}
		\label{fig_5}
	\end{figure}
	
    \begin{figure}[!htb]
		\centering	
		\includegraphics[width=1\columnwidth]{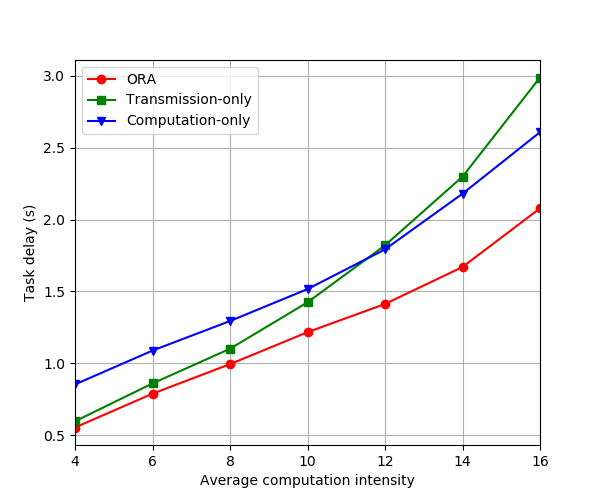}
		\caption{Task delay vs. average computation intensity of tasks.}
		\label{fig_6}
	\end{figure}

Fig. \ref{fig_1} shows how the task delay changes in different epochs. After leaning for a certain number of epochs, the performance does become relatively stable. Meanwhile, we have investigated the impact of the total number of tasks on the average task delay. As shown in Fig. \ref{fig_2}, with the increase of the number of tasks, the task delay of all these three algorithms also increase accordingly and ORA yields a much lower task delay than two other algorithms. It can be observed that ORA reduces the task delay by 14 $\%$ as compared to two other baseline algorithms, when the number of tasks is set to be 600. It is attributed to the fact that ORA can learn to dynamically allocate both the radio and computation resources to each task in real time, it can provision more resources to the current task without significantly degrading the delay of future tasks. In contrast, two other algorithms cannot provision sufficient resources for other tasks after allocating too many resources to the current tasks, thus degrading the average delay of all tasks.

We further investigate the impact of the total number of tasks on the average transmission delay. Fig. \ref{fig_3} shows that the designed algorithm has lower transmission delay than two other algorithms, as the number of tasks increases. Meanwhile, the Transmission-only algorithm has a lower delay than the Computation-only algorithm. As ORA dynamically allocates resources to each task based on the data sizes of tasks and the remaining ratio and computation resource without significantly devastating the performance of the future tasks, it can provision low delay service for tasks. As we know, Transmission-only dynamically allocates radio resources to tasks while provisioning the fixed computation resource for each task, and thus the computing delay becomes a bottleneck. Thus, it has to allocate much more radio resources for some tasks with high computing delay to impose its task delay to meet the QoS constraint, which directly sacrifices the remaining radio resources for other tasks. Therefore, the transmission delay of Transmission-only is higher than that of ORA. On the other hand, while Transmission-only dynamically allocates radio resource to tasks based on their channel conditions and data sizes, Computation-only offers fixed radio resources to tasks and thus incurs a higher transmission delay.  

We also study the impact of the total number of tasks on the average computing delay. Fig. \ref{fig_4} shows that computing delay of ORA is much lower than those of other algorithms when the total number of tasks changes. It is attributed to the fact that ORA considers the current state information such as the channel condition of the IoT device, the data size and computation size of the arriving task, the available radio resource and computation resource of the system. Thus, it can dynamically and fully utilize radio and computation resources to reduce the transmission delay and computing delay. In contrast, the computation resource allocation of Computation-only is affected by its high transmission delay because some tasks with high transmission delay must be allocated with more computation resources to satisfy their QoS requirements. For Transmission-only, since all tasks have the fixed computation resource, it has a higher computing delay than ORA. In addition, we can see that Computation-only has lower computation delay than Transmission-only when the number of tasks is small, and then its computation delay degrades gradually when the number of tasks increases. With small workload, the system have sufficient radio and computation resources for all tasks, and thus Computation-only can dynamically allocates more computation resources to different tasks based on their computation sizes while Transmission-only allocates a fixed computation resource to each task. However, when the workload increases, the performance of Computation-only becomes worse than that of Transmission-only. This is because the tasks in Computation-only are constrained by their fixed radio resources even if they have good channel conditions, and thus incurs a high transmission delay. In this case, Computation-only needs to allocate much more computation resources to these tasks to meet their QoS requirement, thus the remaining computation resources for other tasks are insufficient. As a result, Computation-only has a higher computation delay than Transmission-only when the workload becomes heavy.          

As shown in Fig. \ref{fig_5}, we have studied the impact of average data size of tasks on task delay. It can be seen that the task delay of all these algorithms increases when the average data size increases given the number of tasks ($\lvert \mathcal{I} \rvert$=500). Meanwhile, ORA always has a significantly lower task delay than two other algorithms. It is attributed to the fact that ORA can dynamically adjust the radio and computation resource allocation when the average data size increases, and thus keeps a lower task delay as compared to other algorithms. In contrast, when the average data size increases, the transmission delay becomes a bottleneck for Computation-only while the computing delay is the bottleneck for Transmission-only.  

Fig. \ref{fig_6} illustrates how the task delay changes when the average computation intensity increases. We can see that ORA incurs a significantly lower task delay as compared to two other algorithms. Note that the increase of the average computation intensity impacts the computation sizes of tasks while the data sizes of tasks keep the same. In this case, ORA can learn to adjust the radio and computation resources for different tasks based on their computation sizes and data sizes, and thus incurs a lower task delay than two other algorithms. Furthermore, for a low average computation intensity, the network has much low computation load, and thus the transmission delay becomes the dominating factor of the task delay. In this case, Transmission-only can dynamically allocate the radio resource to tasks and thus incurs a lower task delay than Computation-only in which the radio resource of each task is fixed. However, as the average computation intensity increases, the computation load dramatically increases and thus the computing delay becomes the dominating factor instead. Since Computation-only dynamically allocates computation resources based on tasks' computation sizes, it yields a lower task delay than Transmission-only which allocated fixed computation resource to different tasks.  

\section{Conclusion}
\label{secConclusion}
In this paper, we have designed an online resource allocation algorithm based on reinforcement learning to dynamically allocate resources to IoT tasks to improve the task delay of tasks. As tasks are generated dynamically and the future task information is hard to predicted, the resource allocation for different tasks are coupled with each other. Meanwhile, as the task delay consists of both the transmission delay and computing delay, we have jointly considered the radio and computation resource allocation to improve the task delay of all tasks. Due to the QoS constraint of each task, the radio resource allocation and computation resource allocation are also coupled with each other. The designed algorithm employed actor-critic method to iteratively learn the environment and thus make a right resource allocation decision in real time based on the current state information without the future task information. We have demonstrated the performance of the designed algorithm over other baseline algorithms via extensive simulations.

\bibliographystyle{IEEEtran}
\bibliography{myref}
	
\end{document}